\documentclass[11pt]{article}
\usepackage[utf8]{inputenc}
\usepackage{epsfig}
\usepackage{amsmath,amsfonts,latexsym, amssymb}
\newtheorem{satz}{Theorem}[section]
\newtheorem{defi}[satz]{Definition}

\newtheorem{assumption}[satz]{Assumption}

\newtheorem{conclusion}[satz]{Conclusion}
\newtheorem{ob}[satz]{Observation}

\newtheorem{propo}[satz]{Proposition}

\newcommand{\mcal}{\mathcal}
\newcommand{\mbf}{\mathbf}

\newcommand{\tit}{\textit}

\newcommand{\bewende}{$ \hfill \Box $}

\begin{document}
\thispagestyle{empty}
\begin{center}
\vspace*{1.0cm}

{\LARGE{\bf Spontaneous Symmetry Breaking of\\ Lorentz and (Galilei)
    Boosts in\\ (Relativistic) Many-Body Systems      }} 

\vskip 1.5cm

{\large {\bf Manfred Requardt }} 

\vskip 0.5 cm 

Institut f\"ur Theoretische Physik \\ 
Universit\"at G\"ottingen \\ 
Friedrich-Hund-Platz 1 \\ 
37077 G\"ottingen \quad Germany\\
(E-mail: requardt@theorie.physik.uni-goettingen.de)

\end{center}

\vspace{0.5 cm}

\begin{abstract}
  We extend a result by Ojima on spontaneous symmetry breaking of
  Lorentz boosts in thermal (KMS) states and show that it is in fact a
  special case in a more general class of examples of spontaneous
  symmetry breaking of Lorentz symmetry in relativistic many-body
  systems. Furthermore we analyse the nature of the corresponding
  Goldstone phenomenon and the type of Goldstone excitations (provided
  they have particle character).

\end{abstract} \newpage
\setcounter{page}{1}
%%%%%%%%%%%%%%%%%%%
\section{Introduction}
Spontaneous symmetry breaking (SSB) and the Goldstone phenomenon,
usually accompanying it if interactions are sufficiently short-ranged
and if the broken symmetry is a continuous one, which is for example
the case if it derives from a conserved current, is a widely studied
subject, both in the relativistic and the non-relativistic (typically
many-body) regime. It is here not the place to give an exhaustive list
of references. We rather choose to cite a few papers which treat the
subject in a more rigorous and systematic manner and refer the
interested reader to the respective lists of references. As to
relativistic quantum field theory (RQFT), just to mention a few, we
have e.g. \cite{Ezawa}, \cite{Reeh1}, \cite{Reeh2},
\cite{Requ1}. Adopting a slightly more general point of view, covering
both the relativistic and the non-relativistic regime, one may mention
\cite{Kastler} and \cite{Requ2}. With more emphasis on the field of
non-relativistic many-body physics (ground states and temperature
states) see e.g. \cite{L.Landau}, \cite{Requ3}, \cite{Requ4}.

In most cases the conserved current, being related to the
infinitesimal generator of the continuous symmetry, is
translation-covariant, i.e. there do not occur extra and explicit
space-time dependent prefactors. But there are a few notable
exceptions, for example the currents, being related to the Galilei or
Lorentz boosts. Galilei boosts were, to our knowledge, for the first
time studied in this context in \cite{Swieca} and later in
\cite{Requ5}, a short remark can also be found in
\cite{Requ3}. Somewhat later, when it became fashionable to study also
temperature states in the relativistic regime, a similar analysis was
made for the Lorentz group (\cite{Ojima}).

The infinitesimal generators of Galilei and Lorentz boosts are special
in various respects. First, in the translation-covariant case the
Fourier transform (FTr) of
\begin{equation}(\Omega\,|\,[j_0(\mbf{x},t),A]\,\Omega)\quad\text{or}\quad
  (\Omega\,|\,j_0(\mbf{x},t)\cdot A\,\Omega)     \end{equation}
with respect to $(\mbf{x},t)$ is a measure as has been emphasized and
exploited in some of the above mentioned papers (see
e.g. \cite{Requ3}). This is a useful information as it restricts the
degree of possible singularities in the above FTr (in particular near
$(\omega,\mbf{k})$=(0,0)). In the above expression $j_0$ is the zero
component of the conserved current, the space integral of which (in a
certain sense) defines the infinitesimal generator of the conserved or
broken symmetry. $A$ is an observable which possibly breaks the
symmetry, $\Omega$ the vacuum vector, a ground state or temperature
state in the so-called  KMS- (Kubo-Martin-Schwinger) representation
(see below).\\[0.3cm]
Remark: As we are dealing not exclusively with Minkowski-space, we
sometimes write $(\omega,\mbf{k})$ instead of
$p^{\mu}=(p^0,\mbf{p})$. We will use the two notions
interchangeably.\\[0.3cm]
If explicit prefactors like $x^{\mu}$ occur in the expression of
$j_0(x)$, the above mentioned FTr is no longer a measure but
consists of derivatives of measures which can behave more
singular. Lorentz and Galilei boosts are exactly of this type.

Second, sometimes an important aspect of spontaneous symmetry
breaking (SSB) is overlooked or not really exploited, while it is
crucial for a Goldstone theorem to hold in the strong form. That is,
the (broken) symmetry is usually required to commute with the time
evolution. It is this property which leads in a straightforward way to
the existence of Goldstone modes of vanishing energy for $\mbf{k}\to
0$ (See however the provisos and remarks in section
\ref{Goldstone}). Translation covariance plus locality more or less
guarantee this in the regime of relativistic quantum field theory
(RQFT). In the field of quantum many-body physics, this property is
more subtle (see e.g. \cite{Requ2}, \cite{Requ4}). Neither Galilei nor
Lorentz boosts fulfill this requirement. In order to nevertheless have
a Goldstone phenomenon, some extra discussion is necessary (see
below).

There is another remark to be made as to possible confusions which may
show up if many-body physics and RQFT are merged as in our paper. In
the following we emphasize the fact that this fusion has the effect
that a variety of new excitation modes are typically expected to occur
in the system which are not present in e.g. the vacuum QFT and which
are of a more non-relativistic character. To discern them from the
original (naked) particles, forming the body of the system we are
studying (and being already present in the vacuum QFT), they may be
called \tit{collective excitations} as they are typically excitations
of the whole system. In contrast to that the original particles of the
relativistic vacuum theory become what one usually calls
\tit{quasi-particles}. (We note however, that there is sometimes no
clear distinction of these various modes being made in the
literature). These collective excitations include in particular
specific Goldstone modes, signalling a non-vanishing particle
density. In any case, the corresponding types of SSB are very similar
to corresponding phenomena in the non-relativistic many-body regime,
and hence have been treated in some detail in the corresponding
literature.

On the other hand, one can study the problem what happens for example
to (Goldstone) particles which are already present in the vacuum RQFT
if they are placed in a temperature state. This is an entirely
different question. It is our impression that it is perhaps this
latter row of ideas which provides the motivation for the work in
e.g. \cite{Bros1} or \cite{Bros2}. Be that as it may, in our framework
Goldstone excitations are not! \tit{masked} (cf. \cite{Bros1}). Quite
to the contrary, they usually give a clear signal of their presence in
the form we described already in \cite{Requ3} and in the following. We
note that our findings seem to be in accord with the perturbational
analysis in the non-relativistic framework (see
e.g. \cite{Abrikosov},\cite{Bogoliubov},\cite{Fetter} or the recent
\cite{Arteaga} which actually deals with the relativistic case). It
can of course happen that the excitations are so short-lived (e.g. for
high temperature) so that it make no longer sense to call them particles,
but this phenomenon is also well understood (see e.g. the discussion
in section 6 or in \cite{Requ3}). We think in fact that the seemingly
different approaches are rather complementary and not mutually
exclusive.

The paper is organized as follows. In the next section we introduce
the notion of spectral support of operators, states etc. as a useful
technical tool in the following analysis. We then give a brief account
of the notion of SSB and the Goldstone phenomenon in a more general
context which is followed by a discussion of the particular case of
Lorentz boosts. In sections 5 and 6 we treat the case of temperature
states. We provide arguments that the Goldstone excitations belonging
to the Lorentz boosts (or Galilei boosts) are of phonon type (provided
they have particle character at all) in the interacting case. In
section 7 we give a short discussion of the particle-hole picture of
KMS-states. Section 8 deals with the quite subtle situation in
relativistic many-particle ground states. We show that the emergence
of gapless Goldstone excitations even in the presence of a massive
(free) theory implies certain delicate modifications of the usual
framework, leading to a transmutation of the ordinary energy-momentum
spectrum.

%%%%%%%%%%%%%%%%%%%%%%
\section{The Spectral support of Operators}
As to FTr our conventions are the following. 
\begin{equation}f(x)=(2\pi)^{-n/2}\int e^{-ix\cdot p}\cdot
  \hat{f}(p)\,d^np\quad ,\quad f(p)=(2\pi)^{-n/2}\int e^{ix\cdot
    p}\cdot f(x)\,d^nx     \end{equation}
with $n$ the space-time dimension and
\begin{equation}x\cdot p=x^0p^0-\mbf{x}\mbf{p}=t\omega-\mbf{x}\mbf{k}    \end{equation}
(we set for convenience $c=\hbar=1$). In the Hilbert space of the
respective model theory we are discussing, space-time translations are
denoted by
\begin{equation}U(x)=e^{ix^{\mu}P_{\mu}}    \end{equation}
and act on operators as
\begin{equation}\alpha_x(A)=:A(x)=U(x)\cdot A\cdot U^{-1}(x)   \end{equation}
Then the spectral support of an operator (field, observable) is
defined by
\begin{equation}\int A(x)\cdot f(x)\,dx=:\int \hat{A}(p)\cdot \hat{f}(p)\,dp   \end{equation}
or
\begin{equation}\hat{A}(p)=(2\pi)^{-n/2}\int dx\,  e^{-ix\cdot p}\cdot
A(x)\quad , \quad A(x)= (2\pi)^{-n/2}\int dp\, e^{ix\cdot p}\cdot \hat{A}(p)                   \end{equation}
to be understood in the sense of operator-valued distributions. It can
also be defined with respect to states, automorphism groups and the
like. By mathematicians it is frequently called the
Arveson-spectrum. For physicists, who have been using such concepts
for quite some time, it is the energy-momentum content of a field or
observable. It has for example been systematically used in
\cite{Thirring}. Some of its mathematical properties have been
discussed in the nice review by Kastler (\cite{Kastler2}). 

In the case of a unitary group action the physical content becomes
quite transparent by writing
\begin{multline}\int A(x)\cdot f(x)\,dx=\int dx\,f(x)\int\int
  dE_p\,A\,dE_{p'}\cdot e^{ix(p-p')}=\\
(2\pi)^{n/2}\int\int  dE_p\,A\,dE_{p'}\cdot \hat{f}(p-p')=
(2\pi)^{n/2}\int dq\,\hat{f}(q) \int\int
\delta(q-(p-p'))\,dE_p\,A\,dE_{p'}\\
=:\int\hat{f}(q)\hat{A}(q)dq
  \end{multline}
with
\begin{equation}\hat{A}(q):= \int\int
\delta(q-(p-p'))\,dE_p\,A\,dE_{p'}
       \end{equation}

Lorentz boosts, which we will mainly study in the following sections,
act on the spectrum in the following way. As the Minkowski scalar
product is given by $x\cdot y=(x\,|\,\eta\,y)$ with $\eta$ the Minkowski
metric $Diag\, (1,-1-1-1)$ for $n=4$, and the rhs the ordinary scalar
product, we have, because of $\Lambda^T\,\eta\,\Lambda=\eta$:
\begin{equation}\Lambda x\cdot y=(x\,|\,\Lambda^T\eta\, y)=(x\,|\,\eta\Lambda^{-1}y)=x\cdot \Lambda^{-1}y      \end{equation}
Let us denote the automorphism group, implementing the Lorentz boosts
by $\alpha_{\Lambda}$, we hence have 
\begin{equation}\alpha_{\Lambda}\,\hat{A}(q)=(2\pi)^{-n/2}\int
  e^{-iq\cdot x}\cdot A'(\Lambda^{-1}\,x)\,d^nx      \end{equation}
with
\begin{equation} \alpha_{\Lambda}\,A(x)=A'(\Lambda^{-1}\,x)     \end{equation}
where $A'$ is the image of the direct action of a certain representation of
the Lorentz boosts on the operator $A$ (for example some field). This
extra representation does however not change the spectral
content. This yields
\begin{equation}\alpha_{\Lambda}\,\hat{A}(q)=(2\pi)^{-n/2}\int  e^{-i\Lambda^{-1}q\cdot x'}\cdot A'(x')\,d^nx'=\hat{A}(\Lambda^{-1}q)    \end{equation}
hence, as expected, the spectral support is simply shifted under the
action of $\Lambda$.
%%%%%%%%%%%%%%%%%%%%%%
\section{\label{Goldstone}Spontaneous Symmetry Breaking in a Nutshell}
We give a very brief outline of the essentials of SSB. Let $\alpha_t$
denote the time-evolution, acting on the algebra of observables or
fields (we are not too pedantic in this respect), $\gamma_g$ some
symmetry group, $\Omega$ the vacuum state, some other ground state or
a temperature state (in the GNS-representation). We assume that
$\Omega$ is the only state, being invariant under the dynamics, i.e.
\begin{equation}(\Omega\,|\,\alpha_t(A)\, \Omega)=(\Omega\,|\,A\, \Omega)   \end{equation}
Furthermore, we assume that the symmetry group, $\gamma_g$, and the
time evolution, $\alpha_t$, do commute, i.e. 
\begin{equation}(\Omega\,|\,\gamma_g\cdot\alpha_t(A)\, \Omega)=
  (\Omega\,|\,\alpha_t\cdot\gamma_g(A)\, \Omega)   \end{equation}
Remark: As we already mentioned in the introduction, this is not!
fulfilled by the Lorentz boosts.
\begin{defi}The symmetry, $\gamma_g$, is called spontaneously broken,
  if for some $A$ it holds
\begin{equation}(\Omega\,|\,\gamma_g(A)\, \Omega)\neq (\Omega\,|\,A\, \Omega)  \end{equation}
\end{defi}
\begin{ob}\label{SSB}In the case of SSB and uniqueness of $\Omega$ under the time
  evolution, the group, $\gamma_g$, cannot be unitarily represented.
\end{ob}
Proof: We have for some $A$, assuming that $\gamma_g\cdot A=U_g\cdot
A\cdot U^{-1}_g$
\begin{equation}(U^{-1}_g\,\Omega\,|\,A\,U^{-1}_g\,\Omega)\neq (\Omega\,|\,A \Omega)=(\Omega\,|\,\alpha_t(A)\, \Omega)  \end{equation}
On the other hand, the lhs is equal to 
\begin{multline}(\Omega\,|\,\gamma_g(A)\, \Omega)=
  (\Omega\,|\,\alpha_t\cdot\gamma_g(A)\, \Omega)=\\
(\Omega\,|\,\gamma_g\cdot\alpha_t(A)\, \Omega)=(U^{-1}_g\,\Omega\,|\,\alpha_t(A)\,U^{-1}_g\,\Omega)
 \end{multline}
We conclude that $U^{-1}_g\,\Omega$ is also invariant under the
dynamics. The assumed uniqueness however implies that
\begin{equation}U^{-1}_g\,\Omega=\Omega  \end{equation}
hence
\begin{equation}(\Omega\,|\,\gamma_g(A)\, \Omega)=(\Omega\,|\,A\, \Omega) \end{equation}
for all $A$, which is a contradiction because of the assumed
SSB.\bewende\\[0.3cm]
Remark: This result can be extended to certain symmetries, not
commuting with the time evolution. See the next section about the
Lorentz boosts.\vspace{0.3cm}

Frequently, the symmetry group is a continuous group, having, at least
in a formal sense, infinitesimal generators, which generate the
symmetry in the Hilbert space under discussion. For convenience, we
choose a one-parameter subgroup, that is, we deal with a single
generator. In case the symmetry derives from a \tit{conserved
  current}, $j^{\mu}(x)$ with $\partial_{\mu}\,j^{\mu}(x)=0$, the
formal generator has the form
\begin{equation}G:=\int d^{n-1}x\,j^0(\mbf{x},0)= \int d^{n-1}x\,j^0(\mbf{x},t) \end{equation}
i.e., it is (formally) time independent.\\[0.3cm]
Remark: Further possible indices of $j$ are suppressed for the moment
(there do exist, for example, also higher conserved tensor
currents).\vspace{0.3cm}

In the papers, cited in the introduction, the appropriate way of
dealing with such objects is described in more detail. In the
following we are a little bit sloppy by suppressing a further smearing
with a local test function in the time coordinate. An appropriate
definition is then the following:
\begin{equation}d/ds|_{s=0}(\Omega\,|\,\gamma_s(A)\,\Omega)=\lim_{R\to\infty}(\Omega\,|\,[G_R,A]\,\Omega)       \end{equation}
with 
\begin{equation}G_R:=\int d^{n-1}x\, j^0(\mbf{x},0)\cdot f_R(\mbf{x})      \end{equation}
and
\begin{equation}f_R(\mbf{x}):=f(|\mbf{x}|/R)\quad ,\quad
  f(u)=\begin{cases}1\;\text{for}\; |u|\leq 1\\ 0\;\text{for}\; |u|\geq 2 \end{cases}     \end{equation}
and smooth in between. SSB is now expressed via the non-vanishing of
the following expression
\begin{equation}\lim_{R\to\infty}(\Omega\,|\,[G_R,A(t)]\,\Omega)=const\neq 0      \end{equation}
(and being time independent!).\\[0.3cm]
Remark: Remember that the symmetry was assumed to commute with the
time evolution. This is however not sufficient in all cases. In RQFT
we usually assume locality to hold. In this case the conclusion is
correct. In quantum many-body theory certain cluster assumptions for
the occurring two-point functions have to be made which are however
usually fulfilled if interactions are sufficiently short-range. If
these assumptions are violated commutators can become time-dependent
and the ordinary Goldstone phenomenon does no longer hold, As to the
necessary technical details see e.g. \cite{Requ4} section 2 or the
recent book by Strocchi (\cite{Strocchi})\footnote{In chapter 16 of
  the book by Strocchi we found a comment in a footnote concerning our
  paper \cite{Requ3}. It is claimed that in theorem 1 of that paper we
  made some incorrect statements concerning the limit structure of
  functions like $C(\mbf{k},\omega)$ (as to the definition see the
  following observation) in the limit $\mbf{k}\to 0$. We however
  think, our criticized statements are essentially correct and, given
  the usual standards of that time, formulated in a rigorous way. We
  rather think the criticism in \cite{Strocchi} is presumably the
  result of a missunderstanding (for more details see section 6 of the
  present paper, in particular the remarks after observation 6.1).}\vspace{0.3cm}

We see that the FTr of $f_R(\mbf{x})$ is a $\delta$-sequence, converging
to $\delta(\mbf{k})$ for $R\to\infty$. Furthermore, as the limit in
configuration space is time independent, we arive at the following
observation:
\begin{ob}(Goldstone theorem) With $C(\mbf{k},\omega)$ the
  distributional FTr of $(\Omega|[j^0(\mbf{x},0),A(t)]\Omega)$
we have
\begin{equation}\lim_{R\to\infty}\int d^{n-1}k\,C(\mbf{k},\omega)\cdot
\hat{f}_R(\mbf{k})=const\cdot \delta(\omega)     \end{equation}
where $\hat{f}_R(\mbf{k})\to \delta(\mbf{k})$ for $R\to\infty$.
\end{ob}
Analysing the content of this statement, we are led to the conclusion: 
\begin{conclusion}The distribution $C(\mbf{k},\omega)$ necessarily
  contains a \tit{singular} contribution, passing through
  $(\mbf{k},\omega)=(0,0)$. The nature of this singular part has to be
  analysed in the various regimes, mentioned in the introduction. Its
  character is relatively simple in RQFT in the vacuum sector
  (existence of massless Goldstone particles). In quantum many-body
  theory and/or for temperature states the situation is much more
  complex but also more interesting
  (cf. e.g. \cite{Requ3},\cite{Requ4},\cite{Requ2}; most of the
  analysis was already made in \cite{Requ5}).
\end{conclusion}

The following additional observation is perhaps useful. In the
relativistic regime we usually have locality. This implies that
$[j^0(\mbf{x},0),A(t)]$ vanishes for strictly local $A$ and
sufficiently large $\mbf{x}$ (which depends of course on $t$) as the
two operators become space-like separated. It follows that
$(\Omega|[j^0(\mbf{x},0),A(t)]\Omega)$ has a compact support in
$x$-space and hence its FTr is an analytic function in the variable $\mbf{k}$.
%%%%%%%%%%%%%%%%%%%%%%%%%%
\section{\label{boosts}SSB of the Lorentz Boosts}
In contrast to our previous assumption, the automorphisms,
implementing the Lorentz boosts, do not commute with the space-time
translations, i.e.
\begin{equation}\alpha_{\Lambda}\cdot \alpha_x\neq
  \alpha_x\cdot\alpha_{\Lambda}    \end{equation}
with $\Lambda$ a boost (henceforth we use the symbol $\alpha_{\Lambda}$ instead of $\gamma_{\Lambda}$). We will however see, that a weaker condition
does in general hold which turns out to be sufficient. In the
following we will treat systems and states having a non-vanishing
energy-density (or mass-density), like e.g. relativistic equilibrium
states or many-particle ground states. For convenience we assume that
the states are translation invariant (or are invariant under a
suitable subgroup like crystals).

In the framework of RQFT each system has an energy-momentum tensor,
$T_{\nu\mu}(x)$. With $T_{00}(x)$ the energy density, we hence have
\begin{equation}(\Omega\,|\,T_{00}(x)\,\Omega)= const\neq
  0 \end{equation}
We assume the system being at rest relative to an
inertial frame (IF) $S$, i.e., the expectation of the momentum
density, $(\Omega\,|,T_{0i}(x)\,\Omega)$, vanishes. For reasons of
simplicity we now take an IF, $S'$, moving with velocity $v$ in the
negative $x_1$-direction with the coordinate axes of $S,S'$ being
parallel and with $(x_1,t)=0\;\Leftrightarrow\;(x_1',t')=0$. Neglecting
the other, transversal, space coordinates in the following, we have
($c=\hbar=1$) with $\gamma_L:=(1-v^2)^{-1/2}$:
\begin{equation}x_0'= \gamma_L\cdot(x_0+v\cdot x_1)\quad , \quad  x_1'= \gamma_L\cdot(x_1+v\cdot x_0)  \end{equation}
that is $x'=\Lambda\,x$ with
\begin{equation}\Lambda=\begin{pmatrix}\gamma &
    v\cdot\gamma\\v\cdot\gamma & \gamma\end{pmatrix}  \end{equation}

The components of the energy-momentum tensor transform like
\begin{equation}T'_{\nu\mu}(x)=\alpha_{\Lambda}(T_{\nu\mu}(x))=
  (\Lambda^{-1})^{\nu'}_{\nu}\cdot (\Lambda^{-1})^{\mu'}_{\mu}\cdot T_{\nu'\mu'}(\Lambda\,x)     \end{equation}
and the corresponding expectation value in the state $\Omega$:
\begin{multline}(\Omega\,|\,\alpha_{\Lambda}(T_{01}(x))\,\Omega)=\\(\Lambda^{-1})^0_0\cdot(\Lambda^{-1})^0_1\cdot
(\Omega\,|\,T_{00}(\Lambda\,x)\,\Omega)+(\Lambda^{-1})^1_0\cdot(\Lambda^{-1})^1_1\cdot (\Omega\,|\,T_{11}(\Lambda\,x)\,\Omega)=\\
\gamma_L^2\cdot v\cdot ((\Omega\,|\,T_{00}\,\Omega)+ (\Omega\,|\,T_{11}\,\Omega))
\end{multline}
and, by assumption, in the rest system
\begin{equation}(\Omega\,|\,T_{01}\,\Omega)=0     \end{equation}
Remark: The $T_{11}$-contribution represents the possible pressure
work term which can occur in the momentum density under a Lorentz
boost; cf. e.g. the discussion in the recent \cite{RelThermo}. If the
above sum accidentally vanishes in some model, we can use a modified
observable or a rotated reference frame. 
\begin{conclusion}(SSB of Lorentz boosts) \\
\begin{equation}(\Omega\,|\,\alpha_{\Lambda}(T_{01}(x))\,\Omega)=
  \gamma_L^2\cdot v\cdot (\Omega\,|\,(T_{00}+T_{11})\,\Omega)\neq 0= (\Omega\,|\,T_{01}\,\Omega)  \end{equation}
We see that the lhs is independent of $x$ and in particular of
 $t$. Put differently, the lhs is time independent and non-vanishing.
We can hence write
\begin{equation}(\Omega\,|\,\alpha_{\Lambda}(T_{01}(x))\,\Omega)=
  (\Omega\,|\,\alpha_{\Lambda}\cdot\alpha_x(T_{01}(0))\,\Omega)= (\Omega\,|\,\alpha_x\cdot\alpha_{\Lambda}(T_{01}(0))\,\Omega)      \end{equation}
That is, exactly the same proof as in observation \ref{SSB} goes
through. The Lorentz boosts cannot be unitarily represented and there
are gapless excitations (see below).
\end{conclusion}
This generalizes a result of Ojima (\cite{Ojima}) which was stated for
temperature (KMS) states, using a completely different argument (see
the next section).\\[0.3cm]
Remark: That Lorentz or Galilei invariance is broken in infinitely
extended systems with a non-vanishing mass or energy density, is of
course not surprising. What is however interesting is the concrete
technical (and not so obvious) implementation of this fact within the
given theoretical framework and the row of possible consequences. One
should for example remember that e.g. time translation symmetry is a
fairly obvious phenomenon but leads to the deep and important
consequence of energy conservation.\vspace{0.3cm}

It is perhaps noticeable that this time independence can be shown to
hold for arbitrary observables, $A$. The group law for the
Poincar\`{e} group reads:
\begin{equation}(a_1,\Lambda_1)\cdot
  (a_2,\Lambda_2)=(a_1+\Lambda_1\cdot a_2,\Lambda_1\cdot \Lambda_2)
 \end{equation}
and a corresponding relation for the respective automorphism group or
unitary representation. We have in particular that
\begin{equation}(a,\mbf{1})\cdot
  (0,\Lambda)=(a,\Lambda)=(0,\Lambda)\cdot (\Lambda^{-1}\cdot a,\mbf{1})        \end{equation}
and correspondingly for the respective automorphisms. With $a$ a
vector so that $\Lambda^{-1}\cdot a=\hat{t}$ is a vector, pointing in
the time direction, i.e. $a=\Lambda\cdot\hat{t}$, we hence get:

\begin{equation}\alpha_a\cdot\alpha_{\Lambda}=\alpha_{(a,\Lambda)}= \alpha_{\Lambda}\cdot\alpha_{\hat{t}}     \end{equation}
with $\hat{t}=\Lambda^{-1}\cdot a$. With $a$ varying we can reach any
$t$ on the time axis so that we again arrive at
\begin{equation}(\Omega\,|\alpha_{\Lambda}(A(t))\,\Omega)=
  (\Omega\,|\,\alpha_{\Lambda}\cdot\alpha_t(A)\,\Omega)=
  (\Omega\,|\,\alpha_{\Lambda}(A)\,\Omega)     \end{equation}
due to the assumed translation invariance of $\Omega$.

On the level of infinitesimal generators we then have
\begin{equation}d/dv|_{v=0}(\Omega\,|\,\alpha_{\Lambda}(T_{01}(x))\,\Omega)=\lim_{R\to\infty}(\Omega\,|\,[G_R,T_{01}(x)]\,\Omega)=(\Omega\,|\,(T_{00}+T_{11})\,\Omega)    \end{equation}
and being independent of $x$. Or
\begin{equation}\lim_{R\to\infty}\int\,C(\mbf{k},\omega)\cdot\hat{f}_R(\mbf{k})\,d^{n-1}k= (\Omega\,|\,(T_{00}+T_{11})\,\Omega)\cdot\delta(\omega)    \end{equation}
which again shows the existence of gapless Goldstone excitations.

The infinitesimal generator derives from a conserved current
\begin{equation}J^{\mu,\nu\rho}(x):=T^{\mu\nu}(x)\cdot
  x^{\rho}-T^{\mu\rho}(x)\cdot x^{\nu}    \end{equation}
with
\begin{equation}\partial_{\mu}J^{\mu,\nu\rho}(x)=0    \end{equation}
It represents however only the so-called orbital part of the full
expression in case the fields carry non-trivial representations of the
Lorentz group (cf. e.g. \cite{Itzykson}, sect.1.2.2). It is a
conserved quantity because $T^{\mu\nu}(x)$ is conserved and provided
$T^{\mu\nu}(x)$ is symmetric. The generator density for boosts in the
$x$-direction is
\begin{equation}j^0(x)=x^1\cdot T^{00}(x)-x^0\cdot T^{01}(x)    \end{equation}
As $J^{\mu,\nu\rho}(x)$ is conserved 
\begin{equation}J^{\nu\rho}(t):=\int d^{n-1}x\,J^{0,\nu\rho}(\mbf{x},t)    \end{equation}
is (formally) time independent. For $t=0$ it reads
\begin{equation}J^{01}(t)=J^{01}(0)=\int
  d^{n-1}x\,T^{00}(\mbf{x},0)\cdot x^1    \end{equation}

The translation non-covariance has the effect that 
\begin{equation}U_t\cdot J^{\mu,\nu\rho}(\mbf{x},0)\cdot U_t^{-1}\neq J^{\mu,\nu\rho}(\mbf{x},t)    \end{equation}
as the operators $U_t$ act of course only on the operators in
$J^{\mu,\nu\rho}$ and not on the prefactors $x^i$. This implies for
example that
\begin{equation}(\Omega\,|\,[J_R^{01}(0),A(t)]\,\Omega)\neq  (\Omega\,|\,[J_R^{01}(-t),A]\,\Omega)   \end{equation}
We have instead
\begin{equation}U_t\cdot j^0(\mbf{x},0)\cdot U_t^{-1}=x^1\cdot
  T^{00}(\mbf{x},t)=j^0(\mbf{x},t)+x^0\cdot T^{01}(\mbf{x},t)   \end{equation}
and in integrated form
\begin{equation}U_t\cdot J^{01}(0)\cdot U_t^{-1}=J^{01}(t)+x^0\cdot P^1   \end{equation}
with $P^1$ the total momentum operator in the $x$-direction and
$J^{01}(t)=J^{01}(0)$ due to current conservation.

We hence get
\begin{multline}\lim_{R\to\infty} (\Omega\,|\,[G_R(0),A(t)]\,\Omega)=
  \lim_{R\to\infty} (\Omega\,|\,[G_R(-t),A]\,\Omega)-t\cdot
  (\Omega\,|\,[P^1_R,A]\,\Omega)=\\ (\Omega\,|\,[G_R(0),A]\,\Omega)    \end{multline}
as the second contribution vanishes in the limit. It follows
\begin{ob}Due to the time independence of $G(t)=G(0)$ we have
\begin{equation}\lim_{R\to\infty}(\Omega\,|\,[G_R(0),A(t)]\,\Omega)= \lim_{R\to\infty} (\Omega\,|\,[G_R(-t),A]\,\Omega)=\lim_{R\to\infty} (\Omega\,|\,[G_R(0),A]\,\Omega)  \end{equation}
That is, on the level of vacuum expectation values we have an analoguous
result as in the case of a translation covariant current.
\end{ob}
%%%%%%%%%%%%%%%%%%%%%%%%%%%
\section{Lorentz-Invariance for Temperature States}
The paper by Ojima (\cite{Ojima}) deals with SSB of Lorentz boosts in
KMS-states (i.e. Gibbs-states and their thermodynamic limits). In the
case of RQFT such states are still expected to display typical
properties of vacuum states on the operator level, for example,
commutativity of space-like localized observables. On the other hand,
some other properties as e.g. the spectrum condition (joint spectrum
of energy-momentum contained in the forward cone) are typically
lost. This makes these systems particularly interesting.

The KMS-property is typically expressed as follows:
\begin{equation}\omega\, (A(t)\cdot B)=\omega\, (B\cdot A(t+i\beta))       \end{equation}
for $A$ an analytic element so as to allow analytic continuation to
$A(i\beta)$. $\omega$ is a KMS-state at inverse temperature
$\beta$. In the GNS-representation the corresponding Hilbert space
vector is denoted by $\Omega$. In this representation the KMS-property
is expressed as
\begin{equation}(\Omega\,|\,
  B\Delta^{\beta}A\,\Omega)=(\Omega\,|\,AB\,\Omega) \end{equation}
(where we do not discriminate, for convenience, between the elements
of the observable algebra and their Hilbert space representatives),
and
\begin{equation}\Delta^{\beta}=e^{-\beta\hat{H}}       \end{equation}
with $\hat{H}$ the KMS-Hamiltonian, $\hat{H}\Omega=0$. (For technical
details see e.g. \cite{Bratteli} or \cite{Haag}).

As to the interplay of KMS-property and possible symmetries of the
system the following result is typically invoked.
\begin{satz}If a symmetry, $\gamma_g$, leaves the unique KMS-state,
  $\Omega$, invariant, it commutes with the time evolution,
  $\alpha_t$, implemented by
\begin{equation}\alpha_t(A)=\Delta^{it}\cdot A\cdot \Delta^{-it}        \end{equation}
in the GNS-representation. Or for the corresponding operator
implementation:
\begin{equation}\label{KMS}U_g\,\Omega=\Omega\;\Rightarrow\;[U_g,\Delta]=0=[U_g,J]      \end{equation}
with $J$ the Tomita-conjugation (note that $\Delta$ is unbounded, so
that the above statement has to be formulated a little bit more carefully).
\end{satz}
These statements are corolloraries of slightly more general
statements found in e.g. \cite{Takesaki} or \cite{Winnink}. 

In the following we give a short direct proof of the result (for the
convenience of the reader). It contains however a certain technical
subtlety which seems to have been sometimes overlooked in the
literature. We assume that a unitary symmetry, $U$, leaves the
KMS-state invariant, i.e. $U\,\Omega=\Omega$. With the KMS-property as
in formula (\ref{KMS}) and making the following assumption: 
\begin{assumption}We assume that the symmetry maps analytic elements
  into analytic elements, where in this context analyticity has in
  general (e.g. for Lorentz boosts) to be assumed jointly for both
  $\mbf{k}$ and $\omega$!.
\end{assumption}
we have
\begin{multline}(\Omega\,|\,B\,U\,\Delta^{\beta}\,U^{-1}\,A\,\Omega)=
  (\Omega\,|\,\,U^{-1}\,B\,U\,\Delta^{\beta}\,U^{-1}\,A\,U\,\Omega)=\\ 
(\Omega\,|\,\,U^{-1}\,A\,U\,U^{-1}\,B\,U\,\Omega)=(\Omega\,|\,A\,B\,\Omega)=(\Omega\,|\,B\,\Delta^{\beta}\,
A\,\Omega)   \end{multline}
\begin{propo}Under the above assumptions and with the necessary
  technical precautions we have
\begin{equation}U\cdot\Delta^{\beta}\cdot U^{-1}= \Delta^{\beta}\quad
  ,\quad U\cdot J\cdot U^{-1}=J    \end{equation}
\end{propo}
Proof: The first statement follows from the fact that analytic
elements applied to $\Omega$ generate a dense set. The second property
follows from 
\begin{equation}S=J\cdot\Delta^{1/2}\quad , \quad S\,A\,\Omega=A^*\,\Omega     \end{equation}
which yields
\begin{equation}S\,U\,A\,U^{-1}\,\Omega=U\,A^*\,U^{-1}\,\Omega    \end{equation}
or
\begin{equation}S\,U\,A\,\Omega=U\,A^*\,\Omega=U\,S\,A\,\Omega       \end{equation}
hence
\begin{equation}S\,U=U\,S   \end{equation}
which, together with $U\,\Delta=\Delta\,U$ implies
\begin{equation}J\,U=U\,J   \end{equation}
Remark: We recently found that a similar idea was presented in
\cite{Narnhofer}, sect.2 without (as far as we can see) the above
mentioned necessary technical assumption being made. It would be
interesting to investigate which kind of symmetries fulfill or violate
this assumption.\vspace{0.3cm}

We now show the following:
\begin{ob}The Lorentz boosts fulfill the above assumption.
\end{ob}
Proof: We use the results on the spectral support developed in
sect.2. A Lorentz boost simply transforms a point of the spectrum in
the following way
\begin{equation}\Lambda:\,(\omega,\mbf{k})\to \Lambda\cdot (\omega,\mbf{k})      \end{equation}
Analyticity of an element, $A$, means that $\hat{A}(\omega,\mbf{k})$
has a certain decay property with respect to the energy, $\omega$ and
the momentum $\mbf{k}$. This decay property at infinity is evidently
preserved by the above geometric operation, effected by $\Lambda$,
which is essentially a finite shift of the spectral support of
$A$. \bewende

The reasoning of Ojima now is the following. If the symmetry, induced
by the Lorentz boosts, is conserved, implying that it is unitarily
implemented with $U\,\Omega=\Omega$, the above result shows that $U$
has to commute with the time evolution, provided by the
KMS-evolution. On the other hand, the time evolution is part of the
unitary representation of the Poincar\'{e} group. Therefore it cannot
commute with the Lorentz boosts as it is already forbidden
algebraically. Hence we have
\begin{satz}In a KMS-state, representing a pure phase, the Lorentz
  boosts are spontaneously broken.
\end{satz}
We note that our more general result, derived in the preceding
section, comprises this particular result. We have seen that it is
already sufficient that the energy density or some other appropriate
observable does not vanish in the state $\Omega$. 
%%%%%%%%%%%%%%%%%%%%%%%%%%
\section{The Goldstone Theorem for Lorentz Boosts -- the Temperature
  State}
This topic was briefly discussed by Ojima in \cite{Ojima}. As we do
not restrict our investigation to temperature (KMS) states, it appears
to be reasonable to approach this question a little bit more
systematically, in particular as we already made a quite detailed
analysis of the general situation in \cite{Requ3} (and in our
PhD-thesis\cite{Requ5}), which, apparently, Ojima was not aware of.

In the case of KMS-states we have a relative transparent situation in
Fourier-space. We begin with the translation covariant case. With
\begin{equation}J(\mbf{k},\omega):= F.Tr.\;\text{of}\;
  (\Omega\,|\,j^0(\mbf{x},t)\cdot A\,\Omega)\; ,\; C(\mbf{k},\omega) 
:= F.Tr.\;\text{of}\; (\Omega\,|\,[j^0(\mbf{x},t),
A]\,\Omega)  \end{equation}
the Fourier-transformed KMS-condition reads
\begin{equation}C(\mbf{k},\omega)=(1-e^{-\beta\omega})\cdot
  J(\mbf{k},\omega)
  \end{equation}
In \cite{Requ3} or \cite{Requ5} we exhibited the natural two-sidedness
of the Fourier spectrum of $J(\mbf{k},\omega)$ by using the relation: 
\begin{equation}C(\mbf{k},\omega)= J(\mbf{k},\omega)-\overline{J}(-\mbf{k},-\omega)= (1-e^{-\beta\omega})\cdot
  J(\mbf{k},\omega)\end{equation}
which yields the relations
\begin{equation}Re\, J(-\mbf{k},-\omega)= e^{-\beta\omega}\cdot Re\,
  J(\mbf{k},\omega)\quad ,\quad Im \, J(-\mbf{k},-\omega)=- e^{-\beta\omega}\cdot Im\, J(\mbf{k},\omega) \end{equation}
and with $J(\mbf{k},\omega)$ being a measure.

The infinitesimal generator of the Lorentz boosts in the
$x^1$-direction reads (cf. section \ref{boosts})
\begin{equation}j^0(x)=x^1T^{00}(x)-x^0T^{01}(x)       \end{equation}
with $T^{\nu\mu}(x)$ translation covariant. Hence
\begin{equation}J_{\Lambda}(\mbf{k},\omega)=\partial_{k^1} J^{00}(\mbf{k},\omega)-\partial_{\omega} J^{01}(\mbf{k},\omega)  \end{equation}
with $J^{00},J^{01}$ measures.
\begin{ob}$J_{\Lambda}(\mbf{k},\omega)$ consists of terms which are in
  a distributional sense first derivatives of measures.
\end{ob}

The Goldstone theorem was formulated with the help of
$J(\mbf{k},\omega)$ or $C(\mbf{k},\omega)$ in the following way
(cf. section \ref{Goldstone}):
\begin{equation}\lim_{R\to\infty}\int
  C(\mbf{k},\omega)\cdot\hat{f}_R(\mbf{k})\,d^{n-1}k= const\cdot \delta{\omega}\end{equation}
with $const\neq 0$. In our particular case this reads 
\begin{equation}\lim_{R\to\infty}(1-e^{-\beta\omega})\cdot \int J(\mbf{k},\omega)     \cdot\hat{f}_R(\mbf{k})\,d^{n-1}k= const\cdot \delta{\omega}\end{equation}
with
\begin{equation}J(\mbf{k},\omega)= \partial_{k^1} J^{00}(\mbf{k},\omega)-\partial_{\omega} J^{01}(\mbf{k},\omega)     \end{equation}
Performing the above limit sloppily we would get a non-sensical
result, i.e.
\begin{equation}(1-e^{-\beta\omega})\cdot J(\mbf{0},\omega)= const\cdot \delta{\omega}     \end{equation}
or
\begin{equation}J(\mbf{0},\omega)= const\cdot (1-e^{-\beta\omega})^{-1}\cdot \delta{\omega}   \end{equation}
which is not well-defined.

It turns out that it is better to include a $\omega$-integration in the
limit calculation, or, as we said already in section 3, it is
advisable to include an extra time smearing in the original operator
expression. I.e. we will study
\begin{equation}\lim_{R\to\infty} (\Omega\,|\,[Q_R(t),
A]\,\Omega)= \lim_{R\to\infty} \int (1-e^{-\beta\omega})\cdot
J(\mbf{k},\omega)\cdot e^{i\omega t}\cdot \hat{f}_R(\mbf{k})\,d\omega d^{n-1}k=const  \end{equation}
(the limit being independent of $t$!)

We will now analyse the singularity structure of $J(\mbf{k},\omega)$
in the vicinity of $(\mbf{k},\omega)=(\mbf{0},0)$. We made a detailed
(but of course not exhaustive) analysis in \cite{Requ3} and
\cite{Requ5} and isolated a so-called singular contribution in
$J(\mbf{k},\omega)$. More precisely, we showed in \cite{Requ3} that it
is the imaginary part of $J(\mbf{k},\omega)$ which contains the
information of the Goldstone phenomenon and it is exactly this
imaginary part we are mainly dealing with in this context (it may
however be true that we sometimes forgot to make this sufficiently
explicit). In chapter 16 of \cite{Strocchi} it was shown that the real
part contains in the limit $\mbf{k}\to 0$ the derivative of a
$\delta$-function in $\omega$. As we already proved in \cite{Requ3}
that this real part does not! contribute in the limits we are taking,
we think, the criticism in \cite{Strocchi} is a little bit beside the
point, in particular as, in our view, most of what has been said in
chapter 16 of \cite{Strocchi} (and more) can already be found in
\cite{Requ3} and \cite{Requ4}. Furthermore, we constantly emphasized
in our contributions dealing with this topic and in particular in the
criticized theorem 1 that taking limits carelessly in the Goldstone
context is usually dangerous.

Making now a simplyfying model assumption (only to get an idea what is
happening) and assuming that there exists a sharp excitation branch in
the spectrum of $J(\mbf{k},\omega)$, which is however a certain
idealisation, we inferred that $J(\mbf{k},\omega)$ contains a
contribution of the form
\begin{equation}J_S(\mbf{k},\omega)=J(\mbf{k})\delta(\omega-\sigma(\mbf{k}))-e^{-\beta|\omega|}J(-\mbf{k})\delta(\omega+\sigma(\mbf{k}))  \end{equation}
with $J(\mbf{k})$ a function which becomes singular in $\mbf{k}=0$,
the first term on the rhs being the \tit{particle-excitation branch} and
the second term being the \tit{hole-excitation branch}. 

Time independence of the above limit allows us to set $t=0$ which
further simplifies the expression. For our Lorentz boost this yields
\begin{equation}const= \lim_{R\to\infty}\int
  (1-e^{-\beta\omega})\cdot\partial_{k^1}J^{00}(\mbf{k},\omega)\cdot\hat{f}_R(\mbf{k})\,d\omega
  d^{n-1}k  \end{equation}
with $J^{00}(\mbf{k},\omega)$ being a measure. In the following we
deal with the contribution having positive $\omega$. We hence have to
analyse the expression
\begin{multline}\int
\lim_{R\to\infty}  (1-e^{-\beta\omega})\cdot\partial_{k^1}(J^{00}(\mbf{k})\cdot\delta(\omega-\sigma(\mbf{k}))\cdot\hat{f}_R(\mbf{k})\,d\omega
  d^{n-1}k =\\\lim_{R\to\infty}
\int d^{n-1}k\,(1-e^{-\beta\sigma(\mbf{k})})\cdot\partial_{k^1}J^{00}(\mbf{k}) \cdot\hat{f}_R(\mbf{k})    \end{multline}
and get again
\begin{equation}\partial_{k^1}J^{00}(\mbf{k})\sim
  \sigma(\mbf{k})^{-1}\quad\text{for}\quad \mbf{k}\to 0   \end{equation}

Ojima in \cite{Ojima} mentioned the possibility of a Goldstone
spectrum without particle structure. This was already discussed in
greater generality in \cite{Requ3} and \cite{Requ5} and is in fact
quite a delicate point. It is difficult to discuss this matter from a
general point of view. Below we are making certain, in our view
reasonable and physically motivated, model assumptions. We however
emphasize that, depending on the particular context, i.e. type of
models, value of thermodynamic parameters as e.g. the temperature,
completely different situations may prevail.

We expect for example, that the existence of an interaction favors the
emergence of relatively long-lived Goldstone excitations (as
e.g. phonons). In the free case (i.e. no interaction) there is no
reason that relatively stable compressional modes should exist. So, in
this situation, we expect that the singular contribution, $J_s$, does
not correspond to typical long-lived excitations. However, in the
appendix of \cite{Requ3} we made an explicit calculation for a free
Bose-gas in the temperature regime where a \tit{Bose-condensate} does
exist. The calculation clearly shows that the density-density
correlation function consists of a normal term, $J_n$, containing no
pronounced particle excitation structure, and what we called a
singular contribution, $J_s$, which exactly displays the sharp
excitation branch of the underlying Bose particles as intermediate
states with exactly the diverging weight function, we predicted for
vanishing energy-momentum. The underlying physical origin of this
divergence is the vanishing of the chemical potential in the presence
of a condensate in the usual expression for the $k$-dependent
occupation number of the Bose particles. In that paper the Bose
particles are associated with the Goldstone particles of the SSB of
the \tit{phase invariance} (Condensation of zero-modes)!

In real many-body systems, excitations usually have a finite
lifetime. The deeper apriori reasons for this phenomenon were brought
to light in \cite{Thirring}. A Goldstone excitation branch having a
finite lifetime was modelled in the above mentioned papers as
\begin{equation}J_S(\mbf{k},\omega)=J(\mbf{k})\cdot\chi (\mbf{k})^{-1}\cdot\phi(\omega-\sigma(\mbf{k})/\chi(\mbf{k}))   \end{equation}
with $\phi(s)$ a smooth function of compact support, normalized so
that $\int\phi(s)\,ds=1$, $\sigma(\mbf{k})$ and $\chi(\mbf{k})$ are
smooth functions for $\mbf{k}\neq \mbf{0}$ with
$\sigma(\mbf{0})=\chi(\mbf{0})=0$, positive elsewhere.
\begin{ob}Note that the $\phi$-term is essentially a $\delta$-sequence
  for $\mbf{k}\to\mbf{0}$ with respect to
  $\omega$. $\chi(\mbf{k})^{-1}$ plays the role of the lifetime of the
  respective excitation with energy $\sigma(\mbf{k})$. We see that by
  construction the lifetime becomes infinite for $\mbf{k}\to\mbf{0}$.
\end{ob}
Proof: see \cite{Requ5} or \cite{Requ3}. The general many-body
philosophy is roughly the following (see e.g. \cite{Abrikosov}). These
elementary excitations do not! correspond to exact statistical (or
micro) states of the system, but represent superpositions (wave
packets) of a large number of such micro states with a narrow spread
in energy. \bewende

In the above mentioned papers we showed that under such conditions we
get a Goldstone phenomenon of essentially the same type as in the case
of a sharp excitation branch. If the energy uncertainty,
$\chi(\mbf{k})$, fulfills
\begin{equation}\chi(\mbf{k})\ll\sigma(\mbf{k})\quad\text{for}\quad \mbf{k}\to 0  \end{equation}
one can speak of a particle-like excitation. If, on the other hand, 
\begin{equation}\chi(\mbf{k})\gg\sigma(\mbf{k})\quad\text{for}\quad \mbf{k}\to 0   \end{equation}
holds, the Goldstone excitation has no longer a particle-character but
rather resembles a resonance.
\begin{ob}Note that in both cases the contribution,
  $J_S(\mbf{k},\omega)$, contracts to a sharp singularity for
  $\mbf{k}\to \mbf{0}$, which we showed, was crucial for the Goldstone
  phenomenon. That is, both situations are logically possible and
  will in fact occur in nature.
\end{ob}
Remark: In the more standard Green's-function approach (see
e.g. \cite{Abrikosov},\cite{Fetter}) the lifetime is given by the
distance of the respective poles from the real axis in the complex
plane (typically the second sheet).\vspace{0.3cm}

The question is now, what is the character of the Goldstone
excitations in the case of the SSB of the Lorentz boosts? Before we
discuss this question in more detail a general remark is in order. In
contrast to the non-relativistic regime, where, typically, the
energy-momentum dispersion law of the constituents naturally passes
through $(\omega,\mbf{k})=(0,\mbf{0})$, the relativistic constituents
frequently have a non-zero mass and a dispersion law of the type 
\begin{equation}\sigma(\mbf{k})=\sqrt{m^2+\mbf{k}^2} \end{equation}
i.e., in most of the cases we should expect a gap in the energy
spectrum. Then the (naive) question poses itself of the consistency of
this observation with the predictions of the Goldstone theorem we
derived in the preceeding sections. Ojima calls this a paradox, which
is perhaps an exaggeration. In the case of temperature states he
provides an explanation using the double-sidedness of the spectrum of
$\hat{H}$ (see the next section). Much more interesting is however the
case of relativistic many-body ground states, to which we come
later. In general the following observation is crucial.
\begin{ob}In contrast to naive expectation (resulting perhaps from
  experience with RQFT in the vacuum sector), the excitation spectrum
  of energy-momentum in relativistic states having a non-vanishing
  energy or particle density is richer compared to the situation in
  ordinary RQFT and contains additional excitations which are entirely
  new and rather of a non-relativistic type (phonons, magnons etc.).
\end{ob} 

Therefore representations, which essentially take only into account
contributions in for example so-called Green's functions coming from
the original particle modes of vacuum field theory, are usually
incomplete to a certain degree. One should in this context mention the
ingenious idea of elementary excitations developed by Landau
(\cite{Landau}; see also the remarks in \cite{Thirring}). According to
this philosophy it may be much more convenient to represent the
Hamiltonian as an assembly of weakly interacting elementary
excitations (a kind of canonical transformation) which can
considerably differ from the type of particles one starts from.

The analysis shows that in general the typical low-lying excitations
in quantum liquids of e.g. Bose-type are phonons (see \cite{Landau}
loc.cit. or \cite{Huang}). For an interacting Bose system Landau
conjectures an excitation branch with a linear dispersion law for
small $\mbf{k}$ which goes over into another law for larger
$\mbf{k}$. As to a more rigorous quantitative treatment in form of
perturbation theory see \cite{Abrikosov}, chapt. 25, in particular
chapt. 25.5. A similar discussion can be found in \cite{Fetter},
chapt. 21. What is particularly remarkable is that the phonon
excitation branch shows up as intermediate states in the
perturbational calculation of the 2-point Green's function of the
original massive Bose particles. Usually one would expect them to
occur as intermediate states in density-density correlations which are
composed of 4 field operators. While these calculations are of course
made for non-relativistic many-body systems, there is no reason to
expect a different situation in the relativistic case. These
theoretical observations are further corroborated experimentally by
measuring the specific heat, $c_V$, which shows a $T^3$-behavior for
small $T$, which is typical for a phonon gas. It is notable in this
respect that the free (non-relativistic) Bose gas shows a behavior
$\sim T^{3/2}$ which is typical for massive particles.

As these collective excitations show up if we are in a state of
non-vanishing particle density, these phonon-type excitations have to be
regarded as the Goldstone modes belonging to the SSB of Galilei or
Lorentz boosts in real (i.e. interacting) quantum many-body systems.
\begin{conclusion}In interacting quantum man-body systems the
  Goldstone modes, belonging to the SSB of Galilei or Lorentz boosts
  are of phonon type.
\end{conclusion}
Remark: For Galilei boosts this was already discussed in \cite{Requ3},
sect. 4 or \cite{Requ5}. Note that in the case of the free Bose gas
with a condensate the Bose particles themselves are also Goldstone
particles of the spontaneously broken phase transformation belonging
to the number operator.\vspace{0.3cm}

In \cite{Requ5} we provided an entirely different (qualitative)
argument why the Goldstone particles should be of phonon type. It goes
as follows. In contrast to most cases of SSB which are usually related
to the occurrence of certain phase transitions, and which depend on
the space dimension, Galilei or Lorentz boosts can evidently be broken
in all space dimensions. Assuming then that the dispersion law is
independent of the space dimension, we can argue in the following
way. Making the simplification of a Goldstone excitation of infinite
lifetime, we know from our preceding analysis that in a KMS-state it
holds
\begin{equation}\partial_k
  J(\mbf{k})\sim\sigma(\mbf{k})^{-1}\quad\text{for}\quad k\to
  0   \end{equation}
On the other hand, $J(\mbf{k})$ has to be a locally integrable
function as $J(\mbf{k},\omega)$ is a measure. Making the further
reasonable assumption that for $\mbf{k}\to 0$ 
\begin{equation}\sigma(\mbf{k})\sim k^{\eta}     \end{equation}
we have 
\begin{equation}\partial_k J(\mbf{k})\sim
  k^{-\eta}\quad\text{and}\quad J(\mbf{k})\sim k^{\eta-1}   \end{equation}
for $k\to\ 0$. We hence can conclude that 
\begin{equation}\eta-1< d\quad\text{or}\quad \eta< d+1   \end{equation}
with $d$ the space dimension. Choosing now $d=1$ we get
\begin{conclusion}We conclude from out above qualitative argument that
  the Goldstone mode belonging to SSB of the Galilei or Lorentz boosts
  is expected to have a dispersion law for small $k$, $\sigma(k)\sim
  k^{\eta}$ with $\eta<2$. If the exponent is an integer (which is the
  case in most examples), the only possibility which remains is a
  linear law, $\sigma(k)\sim k$, i.e. of phonon type.
\end{conclusion}
%%%%%%%%%%%%%%%%%%%
\section{The Particle-Hole Picture of KMS-States}
Ojima explained the occurrence of zero-energy Goldstone excitations in
the presence of a massive free Bose field with the help of the
two-sidedness of the spectrum of the KMS-Hamiltonian. For a free Bose
field in a KMS-state (without condensate) the annihilation part of the
Bose field reads (without giving all the technical prerequisites)
\begin{equation}\psi(f)=\psi_F((1+\rho)^{1/2}f)\otimes 1+1\otimes\psi_F^{\dagger}(\rho^{1/2}f)       \end{equation}
with $\psi(f)$ affiliated with the observable algebra, $\mcal{A}$. The
dual field, affiliated with the commutant, $\mcal{A}'$, reads:
\begin{equation}\tilde{\psi}(f)= \psi_F^{\dagger}(\rho^{1/2}f)\otimes 1+1\otimes \psi_F((1+\rho)^{1/2}f)    \end{equation}
The whole construction is performed over the tensor product of two
Fock spaces. The F.Tr. of $\rho$, i.e. $\rho(\mbf{k})$, is the
occupation density of the Bose particles in thermal equilibrium at
inverse temperature $\beta$.

This representation was (to our knowledge) first given by Araki and
Woods for the free non-relativistic Bose field in \cite{araki}. Some
years later (and seemingly independently) it was rediscovered in the
framework of thermo field theory (cf. e.g. \cite{umezawa} or
\cite{ojima2} and references therein). One should note that such a
tensor product representation is in fact rather natural. Assuming for
convenience that the Hamiltonian of a system (e.g. enclosed in a
finite box) has discrete spectrum, it is well-known that the canonical
equilibrium  state over the observable algebra can be extended to a
vector state in a larger Hilbert space having such a tensor product
structure with the observable algebra, $\mcal{A}$, acting in the one
Hilbert space of the product, the commutant, $\mcal{A}'$ in the other
one. This vector state is however not annihilated by the original
annihilation operators belonging to the Bose field (it is in fact a
\tit{cyclic} and \tit{separating} vector). But a so-called
\tit{Bogoliubov-transformation} leads to the usual Fock-space creation
and annihilation operators which enter in the above formula.

In \cite{Thirring} and in more detail in \cite{Requkms}, it was shown
that a similar structure is present in essence also in the case of a
general interacting field theory over a KMS-state. More specifically,
it was rigorously shown that in an approximative sense one can confirm
the Landau picture of elementary excitations in general KMS-states, in
particular the particle-hole picture. This means for example that the
annihilation part of an interacting quantum field consists
approximately of the weighted superposition of the annihilation of an
underlying (quasi) particle mode and the creation  of a respective
hole contribution  being immersed in the infamous 'Dirac sea'. A
corresponding relation holds for the creation part. Furthermore, the
duality between $\mcal{A}$ and $\mcal{A}'$ is established via
corresponding dual fields.

In e.g. \cite{Requ5} the consequences of this picture for SSB were
analysed in quite some detail. In the free case the full Hamiltonian
reads
\begin{equation}\hat{H}=H_F\otimes 1- 1\otimes H_F     \end{equation}
and a similar relation holds for the full symmetry generator
(corresponding relations hold for general finite (volume) systems and
typically, the thermodynamic-limit KMS-Hamiltonian in general is a
certain limit of such expressions).  One can now employ the rigorous
results about the spectral support (Arveson spectrum) given for
example in \cite{Kastler2} which we briefly arrange in the following
for the convenience of the reader. The following spectral results hold
for a cyclic state, $\Omega$:
\begin{itemize}
\item With $\lambda \in Spec(\hat{H})$ there exists an $A\in\mcal{A}$
  for each neighborhood $V(\lambda)$ s.t. $Spec(A)\subset V(\lambda)$
  and $A\Omega\neq 0$.
\item We have in general
\begin{equation}Spec (A_1\cdot A_2)\subset Spec(A_1)\cup
  Spec(A_2)  \end{equation}
\item If $\Omega$ fulfills a certain plausible cluster property under
  e.g. the space translations, the spectrum is even additive, i.e.
\begin{equation}\lambda_1,\lambda_2\in Spec(\hat{H})\Rightarrow
  \lambda_1+\lambda_2 \in Spec(\hat{H})  \end{equation}
The same holds for the joint spectrum of energy-momentum.
\end{itemize}

All this shows that it is in fact easy to construct low-lying
excitation modes from modes which are a distance away from energy
equals zero. I.e., by using the first and the third statement, we can
get a $(\mbf{k},\omega)\approx (\mbf{0},0)$ by composing two
observables, $A_1,A_2$, with energy-momentum concentrated around
\begin{equation}(\mbf{k},\omega), (-\mbf{k}+\delta(\mbf{k}),-\omega+\delta(\omega))     \end{equation}
i.e., by taking
\begin{equation}A_1\cdot A_2\,\Omega\quad\text{with}\quad
  (\mbf{k}_1,\omega_1)+
  (\mbf{k}_2,\omega_2)=(\delta(\mbf{k}),\delta(\omega))\approx (\mbf{0},0)   \end{equation}
%%%%%%%%%%%%%%%%%%%%%
\section{The Goldstone Theorem for Lorentz-Boosts -- the Many-Body
  Ground State}
As a consequence of the strong implications coming from the
KMS-structure of temperature states, the explanation of the occurrence
of gapless excitations was relatively straightforward. The situation
for relativistic many-body ground states, on the other hand, is
surprisingly subtle. The reason is the following. In the case of
KMS-states the existence of \tit{mirror-excitations}, lying in the
\tit{Dirac-sea}, or, stated less poetically, the natural two-sidedness
of the joint energy-momentum spectrum, makes it quite easy to
construct particle-hole states with arbitrarily small energy or
energy-momentum. This construction is not at our disposal in the case
of ground states, which, by definition, have a one-sided
energy-spectrum. By scanning the accessible literature, we found
practically nothing in this direction we can rely on. So, as this is
apparently sort of uncharted territory and, on the other hand, turns
out to be quite intricate, it would be necessary to treat the subject,
i.e. the case of relativistic ground-state models, in considerably
more detail which we would prefer to do elsewhere in order not to blow
up the present paper to much. In the following we rather give a brief
motivation and try to provide some insights.

In order to better understand the impending problems, we analyse the
massive free Klein-Gordon field. To get a better feeling for the
underlying physics, we do not proceed in the most abstract way (like
e.g. in \cite{araki}) and attempt to construct abstract (infinitely
extended) states over the given observable algebra and try to verify
their properties afterwards. We think it is rather advisable to start
from finite volume systems and perform the thermodynamic limit at
finite fixed density. Furthermore, for the sake of physical intuition,
we use the creation-annihilation-operator framework, as it is done in
most of the more physically oriented treatments. While they are
unbounded we are nevertheless not aware of any real problem stemming
from this fact. On the other hand, the bounded Weyl-operator framework
has certain unphysical features as these Weyl-operators contain an
arbitrary number of particle creation and annihilation operators. In
the finite volume case this is a little bit nasty. This effect becomes
only negligible in the infinite-volume limit.

We begin with some notations. We denote the Fock-vacuum by $|0>$. The
particle dispersion law is $\omega_k=(k^2+m^2)^{1/2}$. The commutation
relations between the annihilation and creation operators read
$[a(\mbf{k}),a^{\dagger}(\mbf{k}')]=\delta(\mbf{k},\mbf{k}')$. Note
that we are in the regime of finite volumes, $V$, with periodic
boundary conditions tacitly assumed. That is, the $\mbf{k}$'s are
actually taken from a certain discrete set , which depends on the
volume $V$. As we are in the regime of Bose-statistics, the normalized
$n$-Boson ground state is denoted by $|n>$. It contains $n$ modes of
energy
\begin{equation}\omega_0=m\quad , \quad  \mbf{k}=0      \end{equation}
It is created by $a^{\dagger}(\mbf{0})$ from the Fock-vacuum:
\begin{equation}(a^{\dagger}(\mbf{0}))^n\,|0>=(n!)^{1/2}\,|n>\quad ,
  \quad <n|n>=1        \end{equation}
with
\begin{equation}a^{\dagger}(\mbf{0})\,|n-1>=n^{1/2}\,|n>\quad , \quad a(0)\,|n>=n^{1/2}\,|n-1>      \end{equation}

From this we see that the $n$-particle ground state has an energy gap 
\begin{equation}E_n=n\cdot \omega_0 \end{equation} with respect to the
Fock vacuum. We start from a $n$-particle system in a box of volume
$V$ (periodic boundary conditions) and density $\rho=n/V$. If we now
try to perform the thermodynamic limit, the ground state energy $E_n$
wanders away towards infinity (in contrast to the non-relativistic
scenario!). If we want to arrive at a definite limit theory with a
well-defined Hamiltonian we have, among other things, to renormalize
the energy. While this seems to be pretty obvious there are
nevertheless quite a few technical problems lurking in the background,
for example the loss of relativistic covariance due to the change of
spectrum.

At zero temperature we have in the limit a model system with an
infinite occupation of the ground state, i.e. what is called a
\tit{condensate}. This is known from non-relativistic many-body theory
(cf. e.g. \cite{Bogoliubov}, \cite{Fetter} or \cite{Abrikosov}; one of the early
references is \cite{Pines}) and the methods to
correctly deal with such a phenomenon should be similar in principle
in the relativistic realm apart from the different energy-momentum
dispersion law which makes the treatment more complex (note that in
the non-relativistic case energy is proportional to $k^2$).

The following observation is crucial.
\begin{ob}$a(0)$ and $a^{\dagger}(0)$ trivially commute with all the
  other $a(\mbf{k})$ and $a^{\dagger}(\mbf{k})$ for $\mbf{k}\neq
  0$. Furthermore it holds
\begin{equation}[a(0)/n^{1/2},a^{\dagger}(0)/n^{1/2}]=n^{-1}\to 0      \end{equation}
for $n\to\infty$. Hence, in the limit, the operators
$a(0)/n^{1/2},a^{\dagger}(0)/n^{1/2}$ commute with all the elements of
the algebra $\mcal{A}$.
\end{ob}

We now construct excitations which remain within the $n$-particle
Hilbert space $\mcal{H}_n$. We define
\begin{equation}b^{\dagger}(\mbf{k}):=a^{\dagger}(\mbf{k})\cdot
  a(0)/n^{1/2}\quad\text{for}\quad  \mbf{k}\neq 0       \end{equation}
and 
\begin{equation}b^{\dagger}(\mbf{k})\,|n>=:|n-1,\mbf{k}>\in  \mcal{H}_n        \end{equation}
The ordinary Hamiltonian reads
\begin{equation}H=\sum_{\mbf{k}}\, \omega_k\cdot a^{\dagger}(\mbf{k})a(\mbf{k})       \end{equation}
We have
\begin{equation}H\,|n-1,\mbf{k}>=(\omega_k+(n-1)\cdot \omega_0)\,|n-1,\mbf{k}> =((\omega_k-\omega_0)+n\cdot\omega_0)\,|n-1,\mbf{k}>     \end{equation}
We now define
\begin{defi}In the $n$-particle Hilbert space we define the
  renormalized Hamiltonian
\begin{equation}K:=H-n\cdot\omega_0     \end{equation}
and get
\begin{equation}K\,|n-1,\mbf{k}>= (\omega_k-\omega_0)\,|n-1,\mbf{k}>   \end{equation}
\end{defi}
\begin{ob}We have in $\mcal{H}_n$ 
\begin{equation}K=H-n\cdot\omega_0=\sum_{\mbf{k}}\,(\omega_k-\omega_0)\cdot a^{\dagger}(\mbf{k})a(\mbf{k})    \end{equation}
as at most $n$ terms in the above sum can contribute in
$\mcal{H}_n$. In general we get
\begin{equation}K=H-\mu\cdot \hat{N}\quad \text{with}\quad \mu:=\omega_0    \end{equation}
and $\hat{N}$ the particle number operator in the finite volume Fock-space.
\end{ob}
We now have
\begin{align}e^{iKt}\cdot b(\mbf{k})\cdot
  e^{-iKt} &=e^{-i(\omega_k-\omega_0)t}\cdot b(\mbf{k})\\ 
e^{iKt}\cdot b^{\dagger}(\mbf{k})\cdot e^{-iKt}
&=e^{i(\omega_k-\omega_0)t}\cdot b^{\dagger}(\mbf{k})    \end{align}
and a corresponding result for the $a^{\dagger}(\mbf{k}),a(\mbf{k})$
if $\mbf{k}\neq 0$. For $\mbf{k}=0$ we have
\begin{align}e^{iKt}\cdot a(0)/n^{1/2}\cdot e^{-iKt} &=a(0)/n^{1/2}\\ 
e^{iKt}\cdot a^{\dagger}(0)/n^{1/2}\cdot e^{-iKt}
&=a^{\dagger}(0)/n^{1/2}    \end{align}
We see that in the thermodynamic limit we can treat the zero modes as
(singular!) c-numbers with 
\begin{equation}\lim_{n\to\infty}a_0/n^{1/2}=e^{i\alpha}\cdot \delta{k}    \end{equation}
More specifically, if we smear the respective fields with a test
function $\hat{f}(\mbf{k})$, the zero-component becomes
$\hat{f}(0)\cdot e^{i\alpha}\cdot\mbf{1}$. As to the particular phase
factor see \cite{Haag2} or \cite{araki}. It comes from the additional
breaking of gauge invariance due to the existence of a condensate.

Along these lines one can proceed further, the main task being to cope
properly with the many subtle effects coming from the now missing
relativistic covariance induced by the change in the spectrum of the
Hamiltonian. Note for example that the corresponding redefined quantum
field, based on the $b(\mbf{k})^{(\dagger)}$, does no longer fulfill
the ordinary Klein-Gordon equation but rather some kind of
pseudo-differential equation. In a sense one can however reconstruct
the original covariant Klein-Gordon field. We want to postpone such
an analysis for the reasons mentioned above and instead indicate the
consequences for the Goldstone phenomenon, provided a satisfactory
limit theory can be constructed.

Crucial in this respect are the excitations coming from
$b^{\dagger}(\mbf{k})$ applied to the finite volume $n$-particle
ground state. We saw that $b^{\dagger}(\mbf{k})$ create gapless
excitations for $k\to 0$ with respect to the redefined Hamiltonian
$K$,
\begin{equation}K\cdot
  b^{\dagger}(\mbf{k})\,|n>=(\omega_k-\omega_0)\cdot
  b^{\dagger}(\mbf{k})\,|n>  \end{equation}
with 
\begin{equation}(\omega_k-\omega_0)\approx 1/2m\cdot k^2\quad
  \text{for}\quad k\to 0      \end{equation}

In order to show that the joint energy-momentum spectrum in the
thermodynamic limit is expected to cover the full half-space of
positve energies and arbitrary momenta, we study particular
$N$-particle excitations. We take $N$ excitations,
$(\omega_{k_i}-\omega_0)$, with $|\mbf{k}_i|=\varepsilon_i$. We get
approximately for $\varepsilon_i$ small:
\begin{equation}E_N=\sum_{i=1}^N\,(\omega_{k_i}-\omega_0)\approx 1/2m \sum_{i=1}^N\,\varepsilon_i^2      \end{equation}
Choosing now $\sum_{i=1}^N\,\varepsilon_i=k$ fixed but arbitrary with 
\begin{equation}\varepsilon\geq\varepsilon_i\geq \varepsilon/4\quad
  ,\quad N\cdot\varepsilon/2=k     \end{equation}
we have the simple estimate
\begin{equation}\sum_{i=1}^N\,\varepsilon_i^2\leq
  N\cdot\varepsilon^2=N^{-1}\cdot (2k)^2   \end{equation}
\begin{ob}We can arrange the $\mbf{k}_i$ in such a way that for $N$
  large 
\begin{equation}\sum_{i=1}^N\,\mbf{k}_i= \mbf{k}\quad \text{but}\quad
  E_N\to 0  \end{equation}
\end{ob}
Remark: In a slightly different context the interrelation of the form
of the energy-momentum spectrum and broken Lorentz symmetry is
discussed in \cite{Borchers}.

%%%%%%%%%%%%%%%%%%%
\section{Commentary}
We have seen that the SSB of the Lorentz-boosts (and the
Galilei-boosts) is the consequence of the non-vanishing of the energy
or particle density or perhaps some other density. In so far our
results generalize the result of Ojima which exploited some particular
property of KMS-states. We analyzed in quite some detail the nature of
the Goldstone excitations in the various regimes and showed that in
the interacting case they are of phonon-type. What is remarkable in
this context is the transmutation of the original energy-momentum
spectrum into a gapless spectrum. The gapless excitations are of
\tit{particle-hole type}. In the case of relativistic many-body ground
states the emergence of gapless excitations is particularly subtle.


\begin{thebibliography}{99}
  {\small
\bibitem{Ezawa}H.Ezawa,J.A.Swieca: ``Spontaneous Breakdown of
  Symmetries and Zero-Mass States'', Comm.Math.Phys. 5(1967)330
\bibitem{Reeh1}H.Reeh: ``Symmetry Operations and Spontaneously Broken
  Symmetries in Relativistic Quantum Field Theory'',
  Fortschr.Phys. 16(1968)687
\bibitem{Reeh2}H.Reeh: ``Symmetries, Currents and Infinitesimal Generators'' in Statistical Mechanics and Field
Theory, eds. R.N.Sen,C.Weil, Haifa 1972
\bibitem{Requ1}M.Requardt: ``Symmetry Conservation and Integrals over
  Local Charge Densities in Quantum Field Theory'',
  Comm.Math.Phys. 50(1976)259
\bibitem{Kastler}D.Kastler,D.W.Robinson,J.A.Swieca: ``Conserved
  Currents and Associated Symmetries'', Comm.Math.Phys. 2(1966)108
\bibitem{Requ2}M.Requardt,W.F.Wreszinski: ``Temperature States, Ground
  States, and Relativistic Vacuum States in the Context of Symmetry
  Breakdown'', J.Phys.A: Math.Gen. 18(1985)705
\bibitem{L.Landau}L.Landau,J.Fernando Perez, W.F.Wreszinski: ``Energy
  Gap, Clustering, and the Goldstone Theorem in Statistical
  Mechanics'', J.Stat.Phys. 26(1981)755
\bibitem{Requ3}M.Requardt: ``Dynamical Cluster Properties in the
  Quantum Statistical Mechanics of Phase Transitions'',
  J.Phys.A:Math.Gen. 13(1980)1769
\bibitem{Requ4}M.Requardt: ``About the Poor Decay of Certain
  Cross-Correlation Functions in the Statistical Mechanics of Phase
  Transitions'', J.Stat.Phys. 29(1982)117 
\bibitem{Swieca}J.A.Swieca: ``Range of Forces and Broken Symmetries in
  Many-Body Systems'', Comm.Math.Phys. 4(1967)1
\bibitem{Requ5}M.Requardt: ``Spontane Symmetriebrechung und Phasenuebergaenge in
  der Nichtrelativistischen Vielteilchenphysik'', doctoral thesis,
  Goettingen 1977
\bibitem{Ojima}I.Ojima: ``Lorentz Invariance vs. Temperature in QFT'',
  Lett.Math.Phys. 11(1986)73
\bibitem{Bros1}J.Bros,D.Buchholz: ``The Unmasking of Thermal Goldstone
  Bosons'', Phys.Rev.D 58(1998)125012, hep-th/9608139
\bibitem{Bros2}J.Bros,D.Buchholz: ``Asymptotic Dynamics of Thermal
  Quantum Fields'', Nucl.Phys.B 627(2002)289
\bibitem{Thirring}H.Narnhofer,M.Requardt.W.Thirring: ``Quasi-Particles
  at Finite Temperature'', Comm.Math.Phys. 92(1983)247
\bibitem{Kastler2}D.Kastler: ``Equilibrium States of Matter'',
  Symposia Mathematica Vol. XX (1976)49
\bibitem{Strocchi}F.Strocchi: ``Symmetry Breaking'', Lecture Notes
  in Physics, Springer, Berlin 2005
\bibitem{RelThermo}M.Requardt: ``Thermodynamics meets Special
  Relativity -- or what is real in Physics?'', arXiv:0801.2639[gr-qc]
\bibitem{Itzykson}C.Itzykson,J.B.Zuber: ``Quantum Field Theory'',
  McGraw-Hill, N.Y. 1980
\bibitem{Bratteli}O.Bratteli,D.Robinson: ``Operator Algebras and
  Quantum Statistical Mechanics'', vols. I,II, Springer, Berlin 1979,1981
\bibitem{Haag}R.Haag: ``Local Quantum Physics'', Springer, Berlin 1996
\bibitem{Takesaki}R.H.Hermann,M.Takesaki: ``States and Automorphism
  Groups of Operator Algebras'', Comm.Math.Phys. 19(1970)142
\bibitem{Winnink}M.Sirugue,M.Winnink: ``Constraints imposed upon a
  State of a System that satisfies the KMS Boundary Condition'',
  Comm.Math.Phys. 19(1970)161
\bibitem{Narnhofer}H.Narnhofer: ``Kommutative Automorphismen und
  Gleichgewichtszustaende'', Act.Phys.Austr. 47(1977)1
\bibitem{Abrikosov}A.A.Abrikosov,L.P.Gorkov,I.E.Dzyaloshinski:
  ``Methods of Quantum Field Theory in Statistical Physics'', Dover,
  N.Y. 1975
\bibitem{Bogoliubov}N.N.Bogoliubov: ``Lectures on Quantum
  Statistics'', vol.I, chapt.5, Gordon and Breach, N.Y.1967
\bibitem{Fetter}A.L.Fetter,J.D.Walecka: ``Quantum Theory of
  Many-Particle Systems'', Dover, N.Y. 2003
\bibitem{Arteaga}D.Arteaga: ``Quasiparticle Excitations in
  Relativistic Quantum Field Theory'', arXiv:0801.4324[hep-ph]
\bibitem{Landau}L.D.Landau,E.M.Lifschitz: ``Statistische Physik'',
  vol.5, chapt. 66ff, Akademie Verlag, Berlin 1966
\bibitem{Huang}K.Huang: ``Statistical Mechanics'' chapts. 12 and 18,
  Wiley,  N.Y. 1963
\bibitem{araki}H.Araki,E.J.Woods: ``Representation of the Canonical
  Commutation Relations'', J.Math.Phys. 4(1963)637
\bibitem{umezawa}H.Umezawa,H.Matsumoto,M.Tachiki: ``Thermo Field
  Dynamics'', North-Holland, Amsterdam 1982
\bibitem{ojima2}I.Ojima: ``Gauge Fields at Finite Temperature -- Thermo
  Field Dynamics and the KMS Condition'', Ann.Phys. 137(1981)1
\bibitem{Requkms}M.Requardt: ``A Structure Theorem of General
  KMS-States with a possible Bearing on the Construction of Creation
  and Annihilation Operators for Collective Excitations and Holes'',
  J.Phys. A:Math.Gen. 18(1985)287 
\bibitem{Pines}N.M.Hugenholtz,D.Pines: ``Ground-State Energy and
  Excitation Spectrum of a System of Interacting Bosons'', Phys.Rev. 116(1959)489
\bibitem{Haag2}R.Haag: ``The Mathematical Structure of the
  BCS-Model''. Nuov.Cim. 25(1962)287
\bibitem{Borchers}H.J.Borchers,D.Buchholz: ``The Energy-Momentum
  Spectrum in Local Field Theories with Broken Lorentz-Symmetry'',
  Comm.Math.Phys. 97(1985)169

}
\end{thebibliography}
\end{document}